\documentclass[preprint,showpacs,preprintnumbers,amsmath,amssymb,superscriptaddress]{revtex4}
\usepackage{graphicx}% Include figure files
\usepackage{dcolumn}% Align table columns on decimal point
\usepackage{bm}% bold math
\usepackage{longtable}

\begin{document}

\title{Finite size effects and symmetry breaking in the evolution of networks of competing Boolean nodes}% Force line breaks with \\
\author{Min Liu }
\affiliation{ Department of Physics,
University of Houston, Houston, Texas 77204-5005, USA }
%\email{mliu3@uh.edu}
\author{Kevin E. Bassler}%
\email{bassler@uh.edu} 
\affiliation{ Department of Physics,
University of Houston, Houston, Texas 77204-5005, USA }
\affiliation{Texas Center for Superconductivity, University of Houston, Houston, Texas 77204-5002, USA}
\date{November 14, 2007}% It is always \today, today,
             %  but any date may be explicitly specified

\begin{abstract}
The effects of the finite size of the network
on the evolutionary dynamics of a Boolean network are
analyzed.
In the model considered, Boolean networks evolve
via a competition between nodes that punishes those in the majority.
Previous studies of the model have found that large networks
evolve to a statistical steady state that is both critical and highly
canalized, and that the evolution of canalization,
which is a form of robustness
found in genetic regulatory networks,
is associated with a particular symmetry of the evolutionary dynamics.
Here it is found that finite size networks evolve in
a fundamentally different way than infinitely large networks do.
The symmetry of the evolutionary dynamics of infinitely large
networks that selects for canalizing Boolean functions
is broken in the evolutionary
dynamics of finite size networks.
In finite size networks there is an additional selection for
input inverting Boolean functions that output a value opposite
to the majority of input values.
These results are revealed through an empirical study of
the model that calculates the frequency of occurrence of the
different possible Boolean functions. Classes of
functions are found to occur with the same frequency. Those classes depend
on the symmetry of the evolutionary dynamics and correspond to
orbits of the relevant symmetry group.
The empirical results match analytic results, determined by utilizing
P\'olya's theorem, for the number
of orbits expected in both finite size and infinitely large networks.
The reason for the symmetry breaking in the evolutionary dynamics
is found to be due to the need for nodes in finite size networks
to behave differently in order to
cooperate so that the system collectively performs
as well as possible.
The results suggest that both finite size effects and
symmetry are important for understanding the evolution of
real-world complex networks, including genetic regulatory networks.
\end{abstract}

\pacs{87.23.Kg, 05.65.+b, 87.14.Gg, 89.75.Hc}% PACS, the Physics and Astronomy
                             % Classification Scheme.
%\keywords{Suggested keywords}%Use showkeys class option if keyword
                              %display desired
\maketitle
%------------------------------------------------------------------
%
%  First part: Introduction
%
%------------------------------------------------------------------

\section{Introduction}\label{Intro}

Boolean networks~\cite{K69,K93,ACK02,Drossel07}
%\cite{ K69, K93,Kauffman2,DP86,DW86,BP97,BP98,ACK02,AB00,Drossel07}
have been studied extensively over the past three decades. They consist of a directed
graph in which the nodes have binary output states that are determined by Boolean functions of the
states of the nodes connected to them with directed in-links.
They have
applications as models of gene regulatory networks as well as
models of physical, social, and economic systems.
As ``coarse-grained'' models of genetic networks
they aim to capture the essential features of the dynamical behavior of the real networks while
simplifying local gene expression to a binary (on/off) state~\cite{lessmore}.
Recent work has demonstrated that,
despite their simplicity,
Boolean networks
can indeed describe many of the important features of the dynamics of biological
genetic circuits~\cite{lessE1, lessE2, SKA05, LGT07}.
For example, it has been shown that a Boolean network model of
the segment polarity gene regulatory networks that control
embryonic segmentation in {\em Drosophila melanogaster}
can reproduce the wild-type gene expression pattern and
ectopic patterns due to various mutants~\cite{lessE1}.

Motivated by the fact that evolution plays a crucial role in forming
the regulatory relations among genes,
a number of models that evolve the structure and dynamics of Boolean networks have been
studied~\cite{KS86, BS97, PBC00, Bornholdt, BS00, SOCRBN, canalization, canalization05, coevolution, SD07, R07,BB07}.
These evolutionary Boolean network (EBN) models generally seek to determine the properties of the
networks that result from the evolutionary mechanism being considered.
For example,
many of the studies have focused on the topology of links of the networks that result from
evolutionary mechanisms that rewire the links.
Other studies have focused on finding evolutionary mechanisms that result in networks
that have dynamics that are robust against various types of perturbations, or that result in
networks that are in a
``critical'' state poised between ordered and ``chaotic'' dynamical behavior.

An example of an EBN is the model of competing Boolean nodes
first introduced in Ref.~\cite{PBC00}, and studied subsequently
in Refs.~\cite{canalization, canalization05}.
In this model the nodes of a Boolean network compete with each other in a
variant of the Minority game~\cite{CZ97}.
The network evolves by changing the Boolean function of the node that loses the game to a new
randomly chosen Boolean function. In the principal variant of the model that has been studied,
only the Boolean functions of the nodes evolve, not the links.
Although, it should be noted that changing the Boolean functions used by
the nodes effectively results in a rewiring of the links~\cite{symmetry}.
In the original paper on the model, it was shown that
the network self-organizes
to a statistically steady, nontrivial critical state with this evolutionary mechanism.
Later it was discovered that the critical state the network evolves to
is highly canalized~\cite{canalization}.
Canalization~\cite{W42} is a type of network robustness known
to exist in genetic regulatory networks~\cite{RL98,Q02,BS03}.
It exists when
certain expression states of a subset of the genes that regulate the expression of a gene
control the expression of the gene regardless of whether or not the other genes that otherwise
affect its expression are being expressed.
In this case, the
states of the subset of regulatory genes that control the expression of a gene are
``canalizing'' inputs to the gene.
Canalization is thought to be an important property of developmental biological systems
because it buffers their evolution, allowing greater underlying variation of the genome and
its regulatory interactions
before some deleterious variation can be expressed phenotypically~\cite{W05}.

The canalized nature of the evolved steady state was demonstrated by preforming an
ensemble of similar simulations~\cite{canalization}. Each simulation in the ensemble
involved a different random network
and began with a different initial condition. After they were run long enough to allow the
networks to evolve to a steady state, the average frequency that
the various possible Boolean
functions occurred was measured and averaged over the ensemble of simulations.
For Boolean networks of size $N=999$ nodes in which each node has
$K=3$ in-links, it was found
that the $256 = 2^{2^K}$ possible Boolean functions of three variables
organize into 14 different classes in which all of the functions in each class occur
with approximately the same frequency. It was then found that the various classes of
functions mostly could be
distinguished by the fraction of canalizing inputs that their functions have.
Moreover, the classes whose functions
have larger fractions of canalizing inputs, that is, the ones that are more
canalizing, occurred with larger frequency.

The reason that there are 14 different classes of $K=3$ Boolean functions is due
to the symmetry properties of the evolutionary dynamics~\cite{symmetry}.
The set of Boolean functions of $K$ inputs maps one-to-one onto the set of configurations
of the $K$-dimensional Ising hypercube in which each of the vertices of the hypercube have a
binary state. The different classes correspond to the group
orbits of the ``Zyklenzeiger'' group~\cite{Harrison} which is the hyper-cubic, or hyper-octahedral
symmetry group combined
with parity. In this case, parity means simultaneously inverting the binary states of each of the vertices.
A group orbit is the set of configurations that map into each other through
applications of a group's symmetry operations.  The number of orbits a group has can
be calculated using P\'olya's theorem,
and using P\'olya's theorem it has been shown that there are 14 different group orbits of
the 3 dimensional Zyklenzeiger group~\cite{symmetry}.
Thus, the function classes correspond to orbits of the symmetry group of the evolutionary
dynamics, and for large networks that group is the 3d Zyklenzeiger group.

In this paper, the effects of the finite size of the network on the
self-organized evolution of networks of competing Boolean nodes are studied.
Other EBN models~\cite{BS00, coevolution} have important finite size effects,
indicating that small networks may behave in a fundamentally different way than large networks.
Most previously reported results on this model have been for
relatively large networks with nearly 1000 nodes.
Here, we find that small networks evolve very differently than large networks.
In particular, it will be shown
that the competition of nodes in very small networks does not
result in the evolution of a canalized network.
Instead, the evolutionary mechanism preferentially selects for
``input inverting" Boolean functions that have output states opposite to
the majority of the inputs they receive.
We will also see empirically that for finite size networks
the $K=3$ Boolean functions have 46 different classes, instead of 14.
Finally, it will be shown that this change from
14 to 46 classes is due to a type of symmetry breaking in the evolutionary
dynamics of finite size networks.  P\'olya's theorem will be used
to show that, indeed, 46 classes are expected from the reduced symmetry of
the finite size evolutionary dynamics.

The remainder of this paper is organized as follows. In Sec.~\ref{model}, the definition
of random Boolean networks (RBNs) and of canalizing Boolean functions is given, a few important properties
of RBNs are noted, and
the algorithm used for the evolution of Boolean networks based on a competition between nodes
is stated. Section~\ref{simulation}
presents empirical results obtained from simulations of finite sized networks.
In Sec.~\ref{polya}, P\'olya's theorem is used to calculate the number of
function classes in finite size networks
based on a group of symmetry operations that is reduced from that which
is relevant to the evolution of infinitely large
networks.
Section~\ref{discussion} explains that the symmetry breaking occurs because
the behavior of nodes in finite size networks must differ from those in infinitely
large networks in order for them to cooperate so that they collectively perform as
efficiently as possible.
Finally, a summary of our results and a discussion
of what
they may imply about
the importance of finite size and symmetry in
the evolutionary dynamics of real-world complex  networks is discussed in
Sec.~\ref{conclusions}.

%------------------------------------------------------------------
%
%  Second part: Models
%
%------------------------------------------------------------------
\section{The model}\label{model}
%------------------------------------------------------------------
%
%  Subsection A.
%
%------------------------------------------------------------------
\subsection{Random Boolean Networks}
\begin{figure}[t]
\includegraphics[scale=0.50]{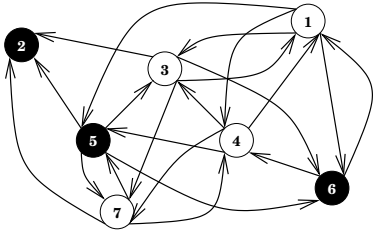}% Here is how to import EPS art
\caption{\label{fig1} An example of a
homogeneous RBN with $N=7$ nodes each of which have in-links from $K=3$ other randomly
chosen nodes. The regulatory directed in-links to a node are indicated by the
arrows pointing to the node on the graph. The black
and white coloring of the nodes indicates the binary output states `1' and `0',
respectively. The
state vector of this network is $\mathbf{\Sigma}(t) = (0,1,0,0,1,1,0)$.}
\end{figure}

In general,
a RBN consists of $N$ nodes, $i=1,\ldots,N$,
each of which has a dynamical Boolean state,
$\sigma_i =0$ or $1$, that is a function of the Boolean states
of $K_i$ other random nodes that regulate its behavior.
The regulatory interactions are therefore described by a directed
graph.
As simple models of genetic networks, nodes can be interpreted as
genes and their Boolean states indicate whether they are
``not being expressed'' or are ``being expressed''.
Different interpretations apply in other cases.
For example, as agent-based models of financial markets,
nodes can be interpreted as traders and their Boolean states
indicate whether they are ``buying" or ``selling".
For synchronously updating RBNs,
which are the only ones considered here,
the state of each node at time $t+1$ is a function of all state of
its $K_i$ regulatory nodes at time $t$.
Therefore, the discrete
dynamics of the network is governed by
\begin{equation}\label{dynamics}
\sigma_i(t+1) = f_i\left(\sigma_{i_1}(t),\sigma_{i_2}(t),\cdots
,\sigma_{i_{K_i}}(t)\right),
\end{equation}
where $i_1$,$i_2$,$\cdots$,$i_{K_i}$ are those $K_i$ nodes that
regulate node $i$. The function $f_i$ is a Boolean function of
$K_i$ inputs that determines the output of node i for all $2^{K_i}$
possible sets of input values.
In this study, for simplicity, we consider only
homogeneous RBNs in which each node has three
input nodes, that is, $K_i = K=3$ for all $i$.
No self-links, or multiple in-links from the same node are allowed.
Figure~\ref{fig1} illustrates an example of the type of homogeneous
RBNs we consider.

Note that random Boolean functions
can be generated with an ``interaction bias" $p$ by setting the
output value for each set of input to be `1' with probability $p$, and
`0' with probability $1-p$. In
modeling gene regulatory networks, the bias $p$ can be interpreted
as a biochemical reaction parameter that defines the probability
that a gene becomes ``active" due to a particular set of regulatory inputs.

Given the Boolean state of each node $i$ at time $t$, $\sigma_i(t)$,
the system state of network is defined as
$\mathbf{\Sigma}(t)=(\sigma_1(t),\cdots,\sigma_N(t))$.
The path
that $\mathbf{\Sigma}(t)$ takes over time $t$ is a dynamical
trajectory in the phase space of system.
Note that there are $2^N$ different system states that are possible.
For synchronous updating of the nodes, because the
dynamics defined in Eq.~\ref{dynamics} is deterministic and the
phase space is finite, all dynamical trajectories eventually
become periodic. That is, after some possible transient behavior,
each trajectory will repeat itself forming a cycle given by
\begin{equation}\label{attractor}
\mathbf{\Sigma}(t) = \mathbf{\Sigma}(t + \Gamma)
\end{equation}
for some time $\Gamma$.
The periodic part of the trajectory is the attractor of the
dynamics, and the minimum $\Gamma > 0$ that satisfies this
equation is the period of the attractor.

The study of the dynamics of RBNs has a long
history~\cite{DP86,DW86, F88, BP97,BP98,Fox,Aldana1,Aldana2, Socolar03,ST03, drossel1,drossel2,drossel3,drossel4, MA05}.
It is known that two distinct phases of dynamical behavior,
``chaotic" and ``ordered", exist for RBNs that have Boolean
functions that are randomly chosen depending on value of the
interaction bias parameter $p$~\cite{DP86,DW86,F88}. For such
homogeneous RBNs with $K>2$, the ``chaotic" phase occurs for values
of $p$ near 0.5, and the ``ordered" phase occurs for values of $p$
near either 0 or 1.
 One way of
distinguishing the two phases is to measure the
distribution of its attractor periods beginning with random initial
states~\cite{BP97}. For RBNs in the chaotic phase the distribution of attractor
periods is sharply peaked near an average value that grows
exponentially with system size $N$, and for RBNs in the ordered
phase the distribution of attractor periods is sharply peaked near
an average value that is nearly independent of $N$.
There is a continuous transition between the two phases,
the so-called ``edge of chaos."
At the phase transition, RBNs are ``critical'' and
have a broad, power-law distribution of attractor
periods.

%------------------------------------------------------------------
%
%  Subsection B.
%
%------------------------------------------------------------------

\subsection{Canalization and Symmetry}

Canalization occurs in Boolean networks when the
Boolean functions that define the dynamics of the nodes are canalyzing.
A Boolean function is canalyzing if its output is fully
 determined by a specific value of one, or more, of its inputs, regardless of
 the value of the other inputs. Such an input value, or set of input values,
 that controls the output of the Boolean function is
 a canalizing input~\cite{K69,K93}.
How canalyzing a Boolean function of $K$ inputs is can be
quantified by a set of
numbers $\mathbf{P}_k, k=0,1,\cdots,K-1$, which are defined the fraction of
the different possible sets of $k$ input values that are canalyzing~\cite{K93}.
Note that for Boolean functions that have $K$ inputs in total, there are
$2^{k} K!/[(K-k)! \; k!]$ different possible sets of $k$ input values.
For example, in a Boolean function with $K=3$ inputs, there are 6 different
possible single ($k=1$) input values, that is, two possible values for each of
the three inputs.

\begin{figure}[t]
\begin{center}
\includegraphics[scale=0.18]{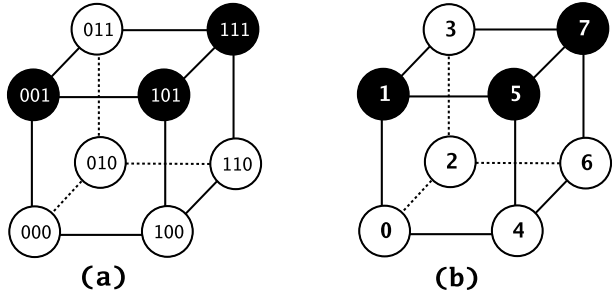}% Here is how to import EPS art
\caption{\label{cube} An example of a $K=3$ Boolean function represented as a
coloring of the Ising cube. The 8 possible sets of input correspond to
the 8 vertices of the cube, and the color of each vertex gives the
output value of the function for that particular set of inputs. Note that
the location of each vertex corresponds to the Boolean vector constructed
from its input values, so that the position of the (0,0,0) vertex is the
origin and the position of the (1,0,1) vertex is $x=1$, $y=0$, and $z=1$.
A white vertex coloring denotes a `0' output value, and, conversely, a black vertex coloring
denotes a `1' output value. In (a) the vertices are labeled by their Boolean
number, and in (b) by their equivalent decimal number.}
\end{center}
\end{figure}

An alternative way to understand canalization can be seen by
mapping the Boolean functions onto configurations, or ``colorings'', of the Ising hypercube~\cite{symmetry}.
Each Boolean
function of $K$ inputs maps one-to-one onto a configuration of
the $K-$dimensional Ising
hypercube. The Ising hypercube is a hypercube which has each vertex labeled, or ``colored'', either `0' or `1'.
In this representation, each of the $2^K$ possible sets of input values corresponds to a different
vertex of the hypercube whose location in $K-$dimensional space is given by the Boolean vector constructed
from its $K$ input values, and the coloring of the vertex corresponds to the output value of the function for that set of
input values.
An example of an Ising hypercube representation for a $K=3$ Boolean function is shown in Fig.~\ref{cube},
where the two possible colors of the vertices, white and black, correspond to the `0' and `1' output values
respectively. Note that the vertices can be labeled either with the binary number constructed from
the sequence of binary inputs, as in Fig.~\ref{cube}(a), or equivalently with the corresponding decimal
number, as in Fig.~\ref{cube}(b). Note that there are 256 different $K=3$ Boolean functions, each corresponding
to a different coloring of the Ising cube.

Note also that each of the $2^{2^K}$ different Boolean functions of $K$ inputs,
or their equivalent colorings of $K-$dimensional Ising hypercubes,
can be uniquely distinguished by an integer $f_D \in [0,2^{2^K}-1]$
defined as
\begin{equation}
f_D = \sum_{i=0}^{2^K-1} 2^i f(i)
\end{equation}
where $f(i)=0$ if vertex $i$ is white and $f(i)=1$ if it is
black. For instance, the Boolean function in Fig.~\ref{cube}
has $f_D=162$.

The
representation of Boolean functions as colorings of Ising hypercubes
can greatly facilitate the recognition of canalizing
functions~\cite{symmetry}. For a $K-$dimensional Ising
hypercube, the fractions of canalizing inputs  $\mathbf{P}_k$ of a Boolean
function are the fraction of its $(K-k)$
dimensional hypersurfaces that are homogeneously colored,
that is, all vertices on the hypersurface have the same color.
In general, there are $2^k K!/[(K-k)! \; k!]$ such hypersurfaces possible.
For example in Fig.~\ref{cube},
the whole cube is not homogeneously colored and so $\mathbf{P}_0=0$,
it has one homogeneously colored face and so $\mathbf{P}_1=1/6$,
and it has $7$
homogeneously colored edges and so $\mathbf{P}_2=7/12$.

With the Ising hypercube representation of Boolean functions it is
easy to see that certain functions are related to others by symmetry.
For example, if you permute the order of the labels of the inputs to a function,
the resulting function and corresponding hypercube coloring will be
related to the original by symmetry. Furthermore, since the ordering of
the input labels is arbitrary the two symmetric functions should be indistinguishable
in the evolutionary dynamics of the network.
This observation suggests
that the symmetry properties of Boolean functions
is important in the dynamics EBNs.
In particular, any symmetry in the dynamics for evolving the Boolean functions
should be
reflected in the symmetry of the evolved steady state, for instance
in the symmetry in the frequency at which the different functions occur.

%------------------------------------------------------------------
%
%  Subsection C.
%
%------------------------------------------------------------------
\subsection{Evolutionary Game}
We consider the following variant of the Minority game for evolving the Boolean functions of a
RBN~\cite{PBC00}.
\begin{enumerate}
\item Begin with an unbiased RBN, that is, one whose functions are randomly chosen such
that all of the different possible Boolean functions are equally likely
to be chosen by each node,
and with
a random initial state.
\item Update the output state of nodes using Eq.~\ref{dynamics},
and determine the attractor of the dynamics~\cite{coevolution}.
\item For each update on the attractor, determine whether `0' or `1'
is the output state of the majority of nodes,
and give a point to each node that is part of the majority.
\item Determine which node is in the majority most often over the length of the attractor.
That is, determine the node with the largest number of points. That node loses the game.
If two, or more, nodes
are tied with the largest score, pick one of them randomly to be the
loser.
\item Replace the function of the losing node
with a new randomly chosen unbiased Boolean function.
\item Return to step 2.
\end{enumerate}
In practice, if the dynamical attractor is longer than some limiting
time, $\Gamma_{\max}$, then score is kept only over that limited time. One iteration
of the game is called an ``epoch". Exactly one evolutionary change
is made to the network each epoch. Note that only the Boolean
functions of the nodes evolve, the topology of the networks
links does not change. The essential features of the
game are (1) frustration~\cite{A94,CZ97}, since most nodes lose each
time step, (2) negative reinforcement, since losing behavior is
punished, and (3) extremal dynamics~\cite{BS93}, since only the worst
performing node's Boolean function is changed. As mentioned above, previous
studies involving large networks with $N=999$ nodes have found that this game
causes the networks to evolve to a critical steady state
that is highly canalized.

%%%%%%%%%%%%%%%
%%
%%%%%%%%%%%%%%%%%%%%%

\section{Simulations and Results}\label{simulation}

Here we report results of simulations of ensembles of different size
networks playing this game of competing Boolean nodes. Again, all of
the networks simulated were homogeneous random networks with $K=3$
inputs per node. The sizes of the networks studied range from
$N=999$ to $N=5$.
Note that $N=5$ is the smallest network with an odd number of nodes
that can be formed without self or multiple connections.
For each $N$ we found that the game caused the
networks to evolve to a statistical steady state, and we measured
the frequency that each of the 256 different functions occurred in
the evolved steady state. This was accomplished by simulating,
for each $N$, an ensemble of
$10^7$ independent realizations. Each realization started with an
independent random network
with random links and different unbiased random functions,
and with an independent random initial state.
The simulation of each realization was run for $10^4$ epochs to allow
the network to evolve to the steady state. At the end of each simulation
the functions used by each node were recorded, and then used to calculate
the average frequency of each function for the ensemble of realizations.

We used the algorithm described in Ref.~\cite{coevolution} to find the dynamical attractor
each epoch.
In searching for the attractor, the maximum attractor period
allowed, $\Gamma_{\max}$, varied with $N$. For $N=999$ it was $10^5$ updates,
and this limit was reached in about half the evolutionary game's epochs.
For all other $N$, $\Gamma_{\max}=10000$. As $N$ decreased, the period limit
was reached in a decreasing fraction of the epochs, until only a small
fraction, about 10\%, reached the limit for $N=19$.
For $N=5$ the maximum possible length of
an attractor is $32$, since $2^N$ is the number of different system states.
Therefore, for networks of this size, the limit
was never reached.
If $\Gamma_{\max}$ is varied from these values, the quantitative results reported below do
change, but the main qualitative findings concerning the symmetry breaking in the dynamics of
finite size networks do not.

\begin{figure}[t]
\begin{center}
\includegraphics[scale=0.50]{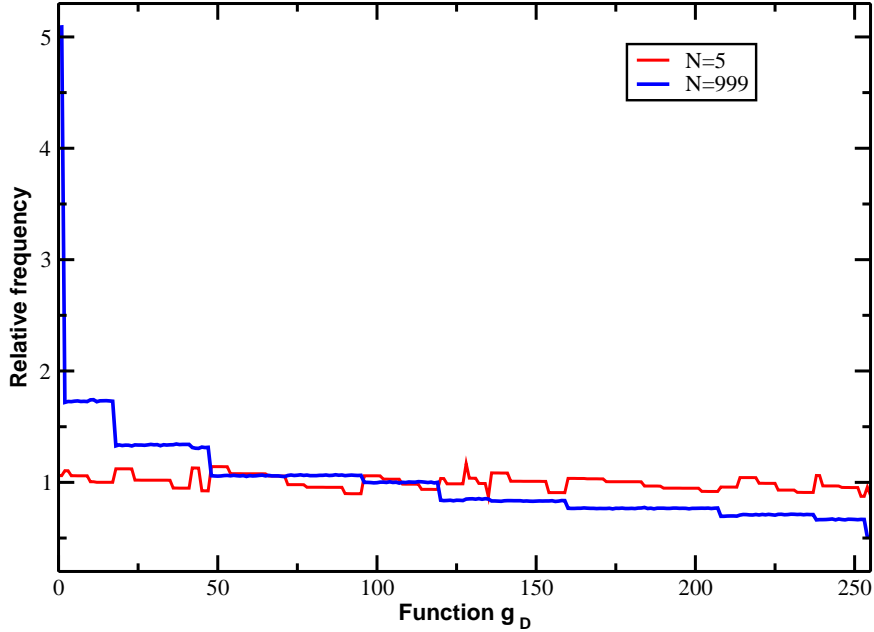}% Here is how to import EPS art
\caption[The relative frequency of occurrence of 256 functions, part a.]
{\label{fig2} (Color online) The ensemble averaged relative frequency at which each of
the 256 $K=3$ Boolean functions occur in the evolved steady state
for networks of size $N=5$ and $N=999$. Networks with $N=999$ appear to have 14 function
classes, while networks with $N=5$ have 46. Note that at the beginning of the simulations
all functions occur with the same relative frequency of 1.0.}
\end{center}
\end{figure}

\begin{figure}[tbh]
\begin{center}
\includegraphics[scale=0.50]{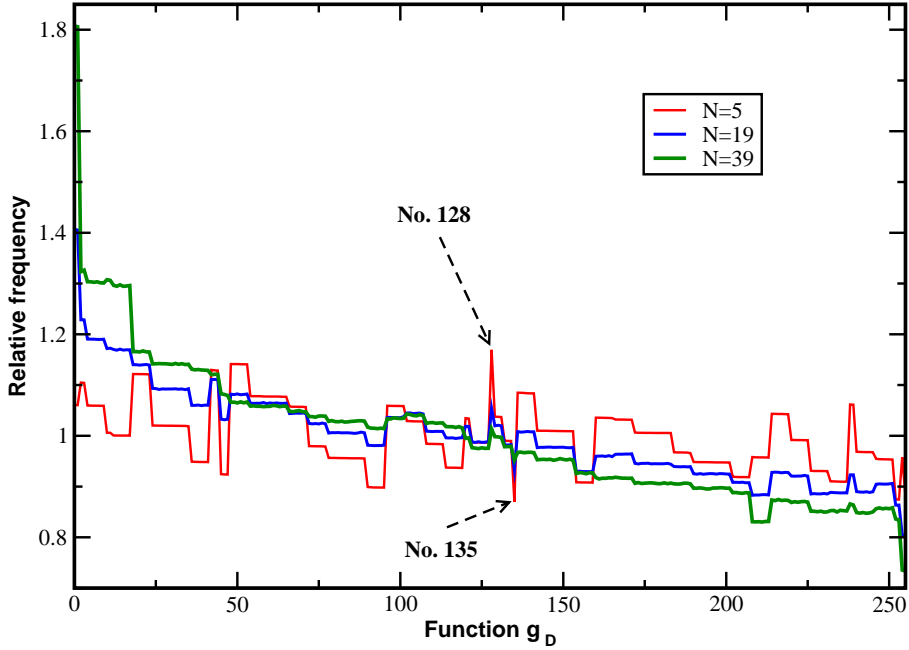}% Here is how to import EPS art
\caption[The relative frequency of occurrence of 256 functions, part b.]
{\label{fig3} (Color online) Ensemble averaged relative frequency at which each of
the 256 $K=3$ Boolean functions occur in the evolved steady state
for networks of size $N=5$, $N=19$ and $N=39$.
These results show that, as $N$ decreases, the evolutionary process shifts from selecting for canalizing
functions to one that selects for input inverting functions.
}
\end{center}
\end{figure}

\begin{figure}[tbh]
\begin{center}
\includegraphics[scale=0.30]{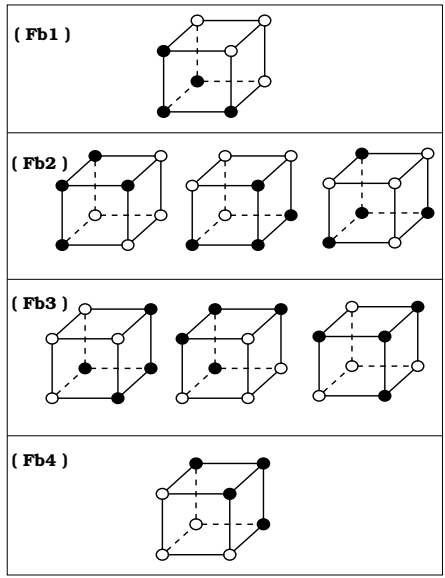}% Here is how to import EPS art
\caption[Illustration of the split of $Fb$ class
consisting of 8 functions.]{\label{breaking}
The Ising cube representations of the 8 Boolean functions, $g_D=128$ through 135, in the $Fb$ class in the
large network limit.  For finite size networks this class splits into 4 classes,
$Fb1$ to $Fb4$, as shown. Functions in each of the four classes all have the same value of
the input inversion factor ${\cal I}$, which are 12, 4, -4, and -12, respectively. }\end{center}
\end{figure}

\begin{figure}[t]
\begin{center}
\includegraphics[scale=0.50]{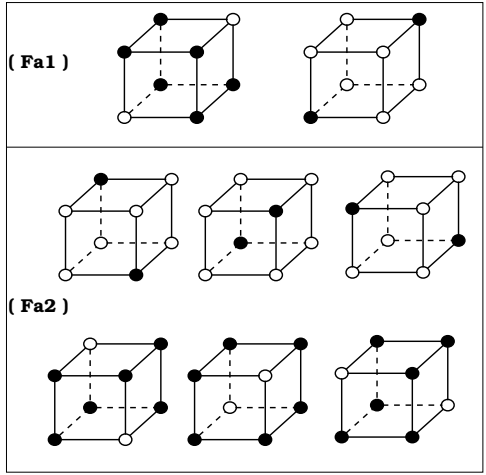}% Here is how to import EPS art
\caption[Illustration of the split of $Fa$ class
consisting of 8 functions.]{\label{Fa_group}
The Ising cube representations of the 8 Boolean functions, $g_D=120$ through 127, in the $Fa$ class in the
large network limit.  For finite size networks this class splits into 2 classes,
$Fa1$ to $Fa2$, as shown. All functions in both classes have the same input inversion factor ${\cal I} = 0.$}
\end{center}
\end{figure}

\begin{figure}[tbh]
\begin{center}
\includegraphics[scale=0.50]{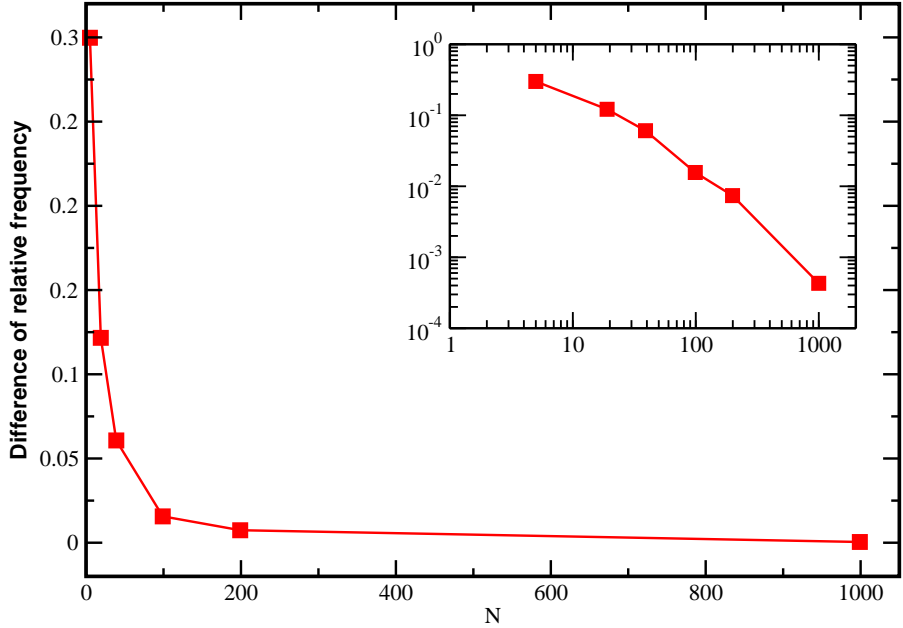}% Here is how to import EPS art
\caption[Effect of size on the canalization.]{\label{fig4}
(Color online) The difference in the relative frequency of functions $g_D=128$ and $g_D=135$
as a function of network size $N$. The inset shows the same data plotted on a log-log scale.
The difference decays approximately as a power law, indicating that the splitting of the function class
vanishes only in the limit of infinite $N$.
}\end{center}
\end{figure}

Figure~\ref{fig2} compares the results for large networks with $N=999$ nodes
with those for small networks with $N=5$ nodes.
Note that all simulations start with an unbiased RBN. Thus at the
beginning of the simulations all Boolean
functions are equally likely
to be chosen by the nodes
and therefore occur with a relative frequency of 1.0.
However, in
the steady state, clearly, some of the 256 possible functions occur more often than others.
Furthermore, there are sets, or classes, of functions that occur with approximately the
same relative frequency.

For the networks with $N=999$ nodes, as observed
previously~\cite{canalization}, there appear to be 14 different classes of
functions.
In order to make this clear, in the figure
the functions numbers have been reordered so that all functions in
the same class are grouped together. Furthermore, all functions in a
class have the same canalization properties and the classes are
sorted in descending order of how canalizing the functions in the
class are. Note that ordered this way, the frequency of the functions
is monotonically decreasing. Therefore, the more canalizing a function
is the more likely it is to occur. The evolutionary process selects
for canalization.

The results are quite different for networks with $N=5$ nodes. In
this case, functions within the 14 classes found for large networks
can now occur with very different frequencies. However, again there
are classes of functions that occur with the same frequency. To see
this note that the functions within each of the 14 classes are
ordered according to the value of their frequency with the functions occurring with
higher frequency first.
Many of the 14 large network function classes
split into multiple classes, and a total of
46 different function classes are observed for small networks.
The complete reordering of the Boolean functions from their decimal numbers
$f_D$ to the numbers $g_D$ used in Figs.~\ref{fig2} and
\ref{fig3}
is given in Appendix~\ref{46classes}.

Note also for $N=5$, that, in contrast to $N=999$, there is not a
monotonic decrease in frequency of the functions with increasing $g_D$,
and therefore the evolutionary process does not select for canalization.
Instead, the evolutionary dynamics selects for  ``input inverting" functions
that output the value opposite to the majority of
their input values.
How input inverting a Boolean function $f$ is can be quantified by its value of
\begin{equation}
{\cal I} = - \sum_{\sigma_1=0}^1 \sum_{\sigma_2=0}^1 \ldots
\sum_{\sigma_K=0}^1 \left[ \left( 2(\sigma_1 + \ldots + \sigma_K) -K
\right)\; \left( 2 f(\sigma_1,\ldots, \sigma_K) - 1 \right) \right],
\label{inversion}
\end{equation}
which is the weighted sum of the number of inverting outputs minus the weighted sum
of the noninverting outputs, where the weight is the strength of the majority in the
inputs. The larger ${\cal I}$ is the more input inverting the function is.
Appendix~\ref{CIIP} lists
the canalization and inversion properties of each class of functions.
Every function in each of the 46 classes has the same value of ${\cal I}$.

The behavior of finite sized networks can be understood in more detail
from the results shown in Fig.~\ref{fig3} which shows the relative frequency of functions
in the evolved steady state for
networks of size $N=5$, 19, and 39. As
network size decreases there is a crossover in the results of the evolutionary
process from selecting for canalizing functions to
selecting for input inverting functions.
For all three sizes of networks, there are 46 classes of functions, and in each
there is combination of preference for canalization and input inversion.
For networks with $N=39$ nodes the most canalizing functions, $g_D = 0$ and 1,
and the least canalizing functions, $g_D = 254$ and 255, occur most often and
least often, respectively.
However,
for networks with $N=5$ nodes the most input inverting function, $g_D = 128$,
and the least input inverting function, $g_D = 135$, occur most often and
least often, respectively.

The splitting of the large network classes can perhaps best be seen by focusing
on the behavior of the functions in the large network class $Fb$, which are
functions $g_{D} = 128$ through 135 in the figure.
The Ising cube representation of these 8 Boolean functions are shown in
Fig.~\ref{breaking}.
Because
${\cal I}=12$ for function $g_D = 128$,
${\cal I}=4$ for functions $g_D = 129$ through 131,
${\cal I}=-4$ for functions $g_D = 132$ through 134,
and ${\cal I}=-12$ for function $g_D = 135$,
the 8 functions split into 4 different classes depending
on the input inversion properties of the functions.

Although the preference for input inverting functions explains much of
the frequency distribution in the evolved steady state, it should be noted
that it does not explain all features of the behavior of small networks. For
example, functions in class $B3$, $g_D = 10$ and 11,
occur slightly more often than those in
class $B4$, $g_D = 12$ through 17,
despite the fact that those in $B4$ are more input inverting.
Also functions in the large network class $Fa$, $g_D=120$ through 127, split into two classes,
$Fa1$ and $Fa2$, even though all functions in $Fa$ are explicitly not input inverting
because their outputs are the same if their input values are simultaneously inverted.
The Ising cube representation of these 8 Boolean functions are shown in
Fig.~\ref{Fa_group}.

The way that the splitting of large network classes quantitatively
depends on network size is shown
in the Fig.~\ref{fig4}, where the difference in the relative
frequency of functions $g_D = 128$ and 135 is plotted as a function of $N$.
The inset shows the same data plotted on a log-log scale.
The frequency difference appears to vanish either as a power law or slightly
faster as $N$ increases.
Analogous behavior occurs for the finite size splitting of the other function classes.
Thus, for any finite value of $N$, even for $N=999$, there is some splitting and there are actually
46 function classes. Only in the limit of infinite $N$ does the splitting vanish
and the number of function classes reduces to 14.

%================
%
%==========================================

%================
%
%==========================================

\section{Symmetry breaking}\label{polya}

Many of the empirical simulation results presented in the previous section
are caused by a breaking of symmetry in the evolutionary dynamics
of finite size networks.
As discussed earlier, each of the 256 Boolean functions of $K=3$ inputs maps one-to-one onto a
coloring of the 3d Ising hypercube, or, equivalently, the Ising cube. The symmetry of the evolutionary
dynamics causes classes of functions to evolve ``symmetrically" with the same frequency.
Thus, the function classes correspond to orbits of the symmetry group of the evolutionary
dynamics~\cite{symmetry}.
In this case,
a group orbit is the set of colorings that map into each other through
applications of the group's symmetry operations.
The number of orbits a group has can
be calculated using P\'olya's theorem~\cite{polya37}.

P\'olya's theorem states that the number of orbits $P_G$ a permutation group $G$ acting on the
$K$-dimensional Ising hypercube has is
$$
P_G
=
\frac{1}{|G|} \sum_{g \in G} |X^g|
$$
where $|G|$ is the number of operators $g \in G$, $X$ is the set of $2^{2^K}$ colorings,
and $X^g$ is the set of colorings that
are left invariant by $g$. To evaluate $|X^g|$, the size of $X^g$,
first express each operator $g$ in terms of its cycle
structure, which is  given by its cycle index
$x_1^{b_1} x_2^{b_2} \cdots x_m^{b_m}$,
where $\sum_{i=1}^m i b_i = 2^K$.
The cycle structure of $g$
describes how it permutes the $2^K$ vertices of the hypercube, and its cycle index
indicates it has $b_1$ cycles of length 1, $b_2$ cycles of length 2, \ldots, and $b_m$ cycles of length $m$.

For example, consider a
2-dimensional square. The four vertices can be labeled 0, 1, 2, and 3
moving clockwise. One permutation $g$ is induced by rotating the square
clockwise through 180 degree. The cycle structure of this operator is $(2
0)(31)$, which means replace 0 by 2 and 2 by 0, and 1 by 3 and 3 by 1. The corresponding
cycle index is $x_2^2$.

For each operator without parity,
the number of colorings left invariant, $|X^g|$, is equal to
$2^{N_c}$, where $N_c=\sum_{i=1}^m b_i$.
Operators with parity must be treated slightly differently.
Parity means simultaneously
reversing the binary state, or color, of
each vertex.
No colorings are left invariant by an operator with parity that has one or more
cycles of length 1, and thus the cycle index of any such operator is defined to be zero.
For each operator with parity,
the number of colorings left invariant, $|X^g|$, is equal to
$2^{N_p}$, where $N_p=(1-\Theta(b_1))\sum_{i=1}^m b_i$ and $\Theta$ is the Heaviside step function.
Note that this means that some operations that include parity may leave no colorings invariant.

Alternatively, the number of orbits, $P_G$, can be calculated by first
finding the ``cycle polynomial" of $G$
by adding the cycle indices for each $g \in G$ together and dividing by $|G|$.
Then, because there are two possible colors of each vertex, replace each $x$ in the cycle
polynomial with 2 and evaluate the expression. The result will be $P_g$.

For infinitely large networks,
the group describing the evolutionary dynamics is the 3d ``Zyklenzeiger" group,
which is the cubic group combined with parity \cite{symmetry}.
The canalization properties of any Boolean function are preserved by the operations of this group.
This group has 96 operations. The operations include the 48 cubic symmetry operations plus each of those
operations combined with parity. The cycle polynomial for this group is
$(1/96)(x_1^8 + 26 x_2^4 + 8x_1^2x_3^2 + 16x_2x_6 + 6x_1^4x_2^2 + 24x_4^2)$,
and the number of group orbits is
$$
P_G
=
\frac{1}{96} \left( 2^8 + 26 \; 2^4 + 8 \; 2^3 + 16 \; 2^2 + 6 \; 2^6 + 24 \; 2^2\right)
=
14.
$$
Note that this is precisely how many function classes are found empirically
in the large network limit.

However, for finite sized networks 46 classes are found empirically. What symmetry
group describes the evolutionary dynamics of finite sized networks?
To answer this question recall  that, as shown in Fig.~\ref{fig3} and discussed
in the previous section, the 46 classes appear from splitting many of the 14 classes that
occur in the large network limit. This indicates that the symmetry group describing
the dynamics of finite size networks is a subgroup of the one for infinitely large networks.
Thus, the symmetry of the dynamics in the large network limit is reduced, or broken, in the dynamics
of finite size networks.

Also recall from the empirical results that when one of the 14 large network classes splits, it
splits into classes that often can be distinguished by
the input inversion properties of the functions.
This was illustrated in Fig.~\ref{breaking} which
shows how the 8 functions in the large network class $Fb$ form four finite size network classes $Fb1$, $Fb2$, $Fb3$, and
$Fb4$ depending on their input inversion properties ${\cal I}$.
While not all of the classes that split from the same large network class can be distinguished
by the value of ${\cal I}$ that their functions have,
all the functions in each of the 46 finite size network classes have the same value of ${\cal I}$.
Thus, the evolutionary dynamics of finite size networks
is consistent with the symmetry of input inversion.

\begin{figure}[tb]
\begin{center}
\includegraphics[scale=0.12]{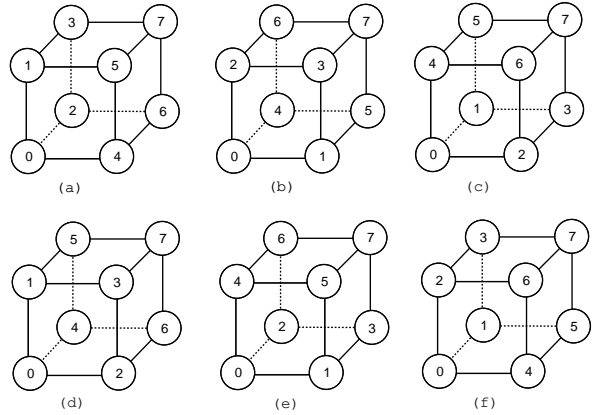}% Here is how to import EPS art
\caption{\label{cubic} Vertex configurations resulting from the 6 symmetry operations
consistent with input inversion that do not involve the parity operation. Six more symmetry
operations are in the group. They are constructed by combining inversion and parity
each of the six operations shown in the figure.}
\end{center}
\end{figure}

However, not all of the symmetry operations in the 3d Zyklenzeiger group that describes
the evolutionary dynamics of infinitely large networks are consistent with
input inversion.
Only those that correspond to changes in the Ising cube that
are symmetric about the (000)-(111) diagonal axis of the cube are consistent with
input inversion.
Thus, of the 96 operations that describe the symmetry of the evolutionary
dynamics of infinitely large networks,
only 12 describe the symmetry of the evolutionary dynamics of finite size networks.
Figure~\ref{cubic} illustrates 6 of these 12 operations by showing the vertex
configuration that
results from the operation.
(a) shows the configuration that results from the identity operation. (b) and (c) show
the configurations that result from rotating
the identity cube 120 and 240 degree counterclockwise about the axis $(000)-(111)$, respectively.
(d) shows the configuration that results from reflecting the identity configuration
about the surface containing
vertices 0, 1, 6, and 7.
Similarly,
(e) and (f) show the configurations that result from
reflecting the identity configuration about the surfaces containing
vertices 0, 2, 5, and 7, and vertices 0, 3, 4, and 7, respectively.
The configurations resulting from the other 6 operations, (g) through (l),
can be obtained by inverting the 6 configurations shown in Fig.~\ref{cubic} about the center of the
cube and applying the parity operation. For example, in (a), inversion means that
exchanging vertex 0 with 7, vertex 3 with 4, vertex 1 with 6, and vertex 2 with 5.

\begin{table}\label{cycletable}
\begin{ruledtabular}
\begin{tabular}[t]{ccc}
Operation $g$ & Cycle Structure & Cycle Index \\ \hline
(a) & $(0)(1)(2)(3)(4)(5)(6)(7)$ & $x_1^8$ \\
(b) & $(0)(241)(653)(7)$ & $x_1^2x_3^2$  \\
(c) & $(0)(421)(563)(7)$ & $x_1^2x_3^2$  \\
(d) & $(0)(1)(42)(53)(6)(7)$ & $x_1^4x_2^2$  \\
(e) & $(0)(41)(2)(63)(5)(7)$ & $x_1^4x_2^2$  \\
(f) & $(0)(21)(3)(4)(65)(7)$ & $x_1^4x_2^2$  \\
(g) & $(07)(61)(52)(43)$ & $x_2^4$  \\
(h) & $(07)(546231)$ & $x_2x_6$  \\
(i) & $(07)(326451)$ & $x_2x_6$  \\
(j) & $(07)(61)(32)(54)$ & $x_2^4$  \\
(k) & $(07)(31)(52)(64)$ & $x_2^4$  \\
(l) & $(07)(51)(62)(43)$ & $x_2^4$
\end{tabular}
\end{ruledtabular}
\caption{The cycle structures and indices of all 12 symmetry operations in the group
consistent with input inversion.}
\end{table}

These 12 operations form a group. This group is input inversion symmetric.
The number of orbits of this group can also be calculated using P\'olya's theorem.
Table~I lists the cycle structure and cycle index for each
of the operations in this group. Adding the cycle indices together, we find that
the cycle polynomial for this group
is
$(1/12)(x_1^8 + 2x_1^2x_3^2 + 3 x_1^4x_2^2 + 2x_2x_6 + 4x_2^4)$,
and that the number of group orbits is
$$
P_G
=
\frac{1}{12} \left( 2^8 + 2 \; 2^4 + 3 \; 2^6 + 2 \; 2^2 + 4 \; 2^4\right)
=
46.
$$
This is exactly what is found empirically. Thus, the input inversion
symmetry group describes the evolutionary dynamics of finite size networks.
The precise form of the evolutionary dynamics is complicated

It is important to note that, although, as discussed earlier,
not all of the classes that split from the same large network class can be distinguished
by the value of ${\cal I}$ that their functions have
and that class $Fa$ splits even though all of the functions in the class are explicitly
not inverting, the selection for input inverting functions by the
evolutionary dynamics is
still responsible for the splitting in these cases.
Because the dynamics of finite size networks selects for input inverting functions, and because
not all of the operations of
the symmetry group for infinitely large networks
are input inversion symmetric,
the symmetry of the evolutionary dynamics of infinitely large networks is
broken.
Once this happens the number of group orbits, or, equivalently, function classes,
increases. Since functions in different classes are not related by the symmetry
of the evolutionary dynamics there is no reason that they should occur with
the same frequency.
It is complicated to calculate the quantitative values of the frequencies of the
functions in the evolved steady state. They depend on more than just their
fractions of canalizing inputs $\mathbf{P}_k$ or their input inversion value ${\cal I}$.
Their frequencies also depend on dynamical interactions of functions
on different nodes. Therefore, in general, it should be expected that unless
functions are in the same orbit of the symmetry group of the evolutionary dynamics
that they not occur with
the same frequency. However, if functions are in the same orbit, then
they are constrained by the symmetry of the evolutionary dynamics to
occur with the same frequency.

\section{Why does the symmetry break?}\label{discussion}
The remaining question to answer is why the evolutionary dynamics of finite size networks
break the symmetry of the evolutionary dynamics of infinitely large networks.
To answer this question note that the evolutionary game the nodes are playing
is a variant of the minority game~\cite{CZ97,PBC00} that effectively punishes nodes when their
output is in the majority and rewards nodes whose output is in the minority.
It has been shown that this effect of the evolutionary dynamics
can result in the
emergence of cooperation and organization of nodes throughout the system
in an attempt to collectively perform as efficiently as possible~\cite{A94,CZ97,PBC00,canalization,LHJ00,HJHJ00,ATBK04}.
The result being that at any time the number of nodes whose output is `0'
is almost equal to, or balanced by, the number of nodes whose output is `1'.
This is true for any size network.
However, the way that this balance is achieved in finite size networks is
different from the way it is achieved in infinitely large networks.

The balance is easier to achieve in large networks than it is in small
networks
because the system has greater flexibility to balance the two states in order to achieve
the global cooperation.
In large networks every one of the $2^{2^K}$ different Boolean functions
is likely to be chosen by some node in every epoch in each realization.
Moreover, any local structure, or motif~\cite{UriAlon,SMMA02}, of the network that
occurs, including the Boolean functions of the associated nodes,
is likely to be simultaneously occurring somewhere else in the network
with opposite parity.
Thus, the balance between nodes with output `0' and nodes with output `1'
can be achieved simply by this ``stochastic" balance that occurs naturally in large
networks.
For infinitely large networks this stochastic balance is exact.

For finite size networks, though, the balance can not be exactly achieved in this
way. There will always be statistical fluctuations that prevent it from happening,
and, for small networks, since each node chooses only one function in any epoch,
there are not enough functions being used for the balance to be achieved stochastically.
The lack of diversity in functions forces the evolutionary dynamics to
place some extra selection pressure on these small number of chosen functions in
order for the system to balance the output of the nodes.
Because of this, it selects for input inverting functions.

To see why input inverting functions help balance the output of the nodes, consider
the extreme case of $N=5$.
In many epochs the attractor that is found has period $\Gamma = 1$.
That is, it is a fixed point, and all nodes always output the same Boolean value.
Clearly, in this case, the best way for a function to work to achieve a collective
balance in the network's output is for it to output the value opposite to the majority of
its input values, that is, for it to be input inverting. If a node's function is
not input inverting, then its output will always be in the majority and the node
may lose the game and have its function changed. Input inverting functions are also
favored in epochs with longer period attractors, but not as strongly.
This mechanism for selecting input inverting functions also works in
larger networks, but becomes decreasingly important as $N$ grows and the nodes
become able to balance the collective behavior stochastically. This
is why, as we found empirically,
the symmetry breaking decreases as $N$ increases, and vanishes in the
limit of infinitely large networks when exact collective balance can be
achieved stochastically.

It is therefore in epochs where the attractor period is short that the selection
for input inverting functions occurs. This contrasts with
how canalization is selected for. It has previously been demonstrated that
canalization is selected for by large networks in epochs in which attractors with
long periods are found~\cite{canalization}. Thus, in the evolved steady state
of finite size networks, since a broad distribution of attractor periods are
found, selection for both canalizing functions and input inverting functions occurs.
However, as we also found empirically,
as $N$ increases and the average length of an attractor period increases, the
selection pressure for input inverting functions decreases while the
selection pressure for canalizing functions increases.

\section{Conclusions}\label{conclusions}

In this study of Boolean networks that evolve because of a
competition between nodes for limited resources, we have found
that the finite size of a network significantly affects
the way that the network evolves and what the emergent properties
of the system are.
In small networks, the evolutionary dynamics selects for input inverting functions,
but as the size of the network increases the evolutionary dynamics
shifts from selecting for input inverting functions to selecting for
canalizing functions until for infinitely large networks only canalization
is selected for. This result can be understood in terms of the
symmetry of the evolutionary dynamics.
For networks in which each node receives 3 inputs,
if the network is infinitely large, then
the evolutionary dynamics has the symmetry of the 3d Zyklenzeiger group, which
is the cubic group combined with parity, but, if the network is of finite size,
then the evolutionary dynamics has the reduced symmetry of the subgroup
consistent with input inversion.
Thus, the symmetry of the evolutionary dynamics of infinite size networks is
broken in the evolutionary dynamics of finite size networks.
The symmetry of the evolutionary dynamics was revealed by the distribution
of the frequency of occurrence of the various Boolean functions, which
shows that there are classes of functions that all occur with the same
frequency, and thus evolve symmetrically. Fourteen function classes are found
for infinitely large networks, while 46 are found for finite size networks.
The function classes correspond to orbits of the symmetry group of the
evolutionary dynamics, and the number of orbits for both of the symmetry
groups
were calculated using P\'olya's theorem and found to match the above
numbers. The reason for the symmetry breaking is that the way the
nodes must behave to collectively perform as well as possible
requires that the evolutionary dynamics be different for infinitely large
and finite size networks, and this difference requires that they select
for different properties in the Boolean functions.

Our previous work on the co-evolution of topology and dynamics of
Boolean networks similarly found
that finite size can affect the evolution of a network.
In that study, which considered a different type of evolutionary dynamics,
we found that finite size networks evolve to have
a broadly distributed in-degree connectivity, whereas infinitely
large networks evolve to have a homogeneous
indegree connectivity~\cite{coevolution}.
Bornholdt and Rohlf also found important finite size effects in
another evolutionary Boolean network model~\cite{Bornholdt}.

As mentioned in the introduction, Boolean networks have been used
to model a wide range of biological, physical, social, and economic
systems, including both gene regulatory networks and financial
markets. Many of these systems evolve through some evolutionary dynamics,
and can be modeled by an EBN.
Although we have focused on just one particular EBN model in this study, our
results show that symmetry and finite size can have important
effects on the evolutionary process.
The properties of the evolved state can
indicate much about the evolutionary process that leds to it.
Real networks, even if they can't be modeled by Boolean networks,
are, of course, always of finite size,
and symmetries exist in many real-world evolutionary  processes.
Therefore, our results suggest that finite size effects and
symmetry may be important in the evolution of the complex networks describing
many real-world systems.

\begin{acknowledgments}
We thank Bogdan Danila for his careful reading of and comments on drafts of this paper.
This work was supported
by the NSF through grant No. DMR-0427538.
\end{acknowledgments}

\appendix
\section{The 46 Function Classes of $K=3$ Boolean functions}\label{46classes}
The classification of the 256 Boolean functions with $K=3$ inputs is
shown in the following table.
The name of each class consists of two parts. The first part is
letters, such as A and Ca, that indicate the
14 classes found in the evolution of infinitely large networks.
The classification scheme was also used in Ref.~\cite{canalization}.
The second part is numbers that indicate the multiple classes that one of the 14
classes split into during the evolution of
finite size networks. The total number of finite size network classes
is 46. For instance,
class $D$ for infinitely large networks splits into 6 classes ($D1$ through $D6$).
The functions are distinguished by both their decimal numbers $f_D$ that uniquely
determine the corresponding coloring of the Ising cube, and
by their reordered numbers $g_D$ that sequentially places functions that are in the same
class.

\begin{longtable}{|c|c|c|}
\hline
Class & Function $g_D$ & Function $f_D$\\
\hline
A   & $0-1$ & 0, 255 \\ \hline
B1  & $2-3$ & 1, 127 \\
B2  & $4-9$ & 223, 4, 16, 247, 2, 191 \\
B3  & $10-11$ & 254,128 \\
B4  & $12-17$ & 239, 251, 32, 8, 253, 64 \\ \hline
Ca1 & $18-23$ & 5, 119, 95,3, 63,17 \\
Ca2 & $24-35$ & 48, 80, 10, 245, 68, 34, 221, 207, 187, 12, 175, 243 \\
Ca3 & $36-41$ & 252, 238, 250, 192, 160, 136\\ \hline
Cb1 & $42-44$ & 15, 85, 51 \\
Cb2 & $45-47$ & 204, 240, 170 \\ \hline
D1  & $48-53$ & 87, 55, 21, 19, 7, 31 \\
D2  & $54-65$ & 117, 69, 47, 81, 35, 13, 79, 59, 93, 11, 49, 115 \\
D3  & $66-71$ & 143, 14, 179, 50, 213, 84 \\
D4  & $72-77$ & 241, 205, 171, 112, 76, 42 \\
D5  & $78-89$ & 138, 140, 186, 174, 220, 206, 208, 244, 162, 176, 196, 242\\
D6  & $90-95$ &  248, 168, 224, 234, 236, 200 \\ \hline
E1  & $96-101$ & 20, 215, 18, 183, 159, 6 \\
E2  & $102-107$ & 65, 111, 9, 123, 33, 125 \\
E3  & $108-113 $ & 130, 144, 246, 132, 190, 222\\
E4  & $114-119$ & 96, 72, 235, 40, 237, 249 \\ \hline
Fa1 & $120-121$ & 129, 126 \\
Fa2 & $122-127$ & 219, 36, 66, 189, 231, 24 \\ \hline
Fb1 & $128$ & 23 \\
Fb2 & $129-131$ & 43, 77, 113 \\
Fb3 & $132-134$ & 142, 178, 212 \\
Fb4 & $135$ & 232 \\ \hline
Fc1 & $136-141$ & 27, 83, 39, 29, 53, 71 \\
Fc2 & $142-153$ & 58, 116, 197, 177, 46, 92, 209, 114, 141, 139, 163, 78 \\
Fc3 & $154-159$ & 184, 172, 202, 216, 228, 226\\ \hline
G1 & $160-165$ & 133, 62, 131, 118, 94, 145 \\
G2 & $166-171$ & 103, 61, 37, 67, 91, 25 \\
G3 & $172-183$ & 181, 211, 70, 199, 26, 52, 157, 38, 28, 155, 167, 82 \\
G4 & $184-189$ & 137, 110, 124, 122, 161, 193 \\
G5 & $190-201$ & 100, 56, 74, 227, 98, 173, 185, 229, 217, 203, 44, 88 \\
G6 & $202-207$ & 218, 230, 194, 188, 152, 164 \\ \hline
Ha & $208-213$ & 102, 60, 195, 90, 153, 165 \\ \hline
Hb1 & $214-219$ & 30, 86, 54, 147, 135, 149 \\
Hb2 & $220-225$ & 45, 99, 101, 89, 57, 75 \\
Hb3 & $226-231$ & 210, 166, 154, 156, 180, 198 \\
Hb4 & $232-237$ &  169, 108, 106, 120, 201, 225 \\ \hline
I1  & $238-239$ & 22, 151 \\
I2  & $240-245$ & 146, 134, 182, 158, 214, 148 \\
I3  & $246-251$ & 121, 109, 41, 107, 73, 97 \\
I4  & $252-253$ & 233, 104 \\ \hline
J1  &  $254$ & 150 \\
J2  &  $255 $& 105 \\
\hline
\end{longtable}

\section{Canalization and Input Inversion Properties of the $K=3$ Boolean Functions}\label{CIIP}
The following table shows the fractions of canalizing inputs $\mathbf{P}_k$ and
the input inversion value ${\cal I}$ for each of the 256 Boolean functions with
$K=3$ inputs. They are grouped according to their 46 classes that occur in the evolution of
finite size networks.

\begin{longtable}{|c|c|c|c|c|c|}
\hline
Class & Function $g_D$ & $\mathbf{P}_0$  & $\mathbf{P}_1$ & $\mathbf{P}_2$ & $\cal I$ \\
\hline
A   & $0-1$   & 1 & 1 & 1 & 0 \\ \hline
B1  & $2-3$   & 0 & $1/2$ & 3/4 & 6 \\
B2  & $4-9$   & 0 & 1/2 & 3/4 & 2  \\
B3  & $10-11$ & 0 & 1/2 & 3/4 & $-6$\\
B4  & $12-17$ & 0 & 1/2 & 3/4 & $-2$\\ \hline
Ca1 & $18-23$ & 0 & 1/3 & 2/3 & 8 \\
Ca2 & $24-35$ & 0 & 1/3 & 2/3 & 0\\
Ca3 & $36-41$ & 0 & 1/3 & 2/3 & $-8$ \\ \hline
Cb1 & $42-44$ & 0 & 1/3 & 2/3 & 8\\
Cb2 & $45-47$ & 0 & 1/3 & 2/3 & $-8$ \\ \hline
D1  & $48-53$ & 0 & 1/6 & 7/12 & 10 \\
D2  & $54-65$ & 0 & 1/6 & 7/12 & 6\\
D3  & $66-71$ & 0 & 1/6 & 7/12 & 2\\
D4  & $72-77$ & 0 & 1/6 & 7/12 & $-2$ \\
D5  & $78-89$ & 0 & 1/6 & 7/12 & $-6$\\
D6  & $90-95$ & 0 & 1/6 & 7/12 & $-10$\\ \hline
E1  & $96-101$  & 0 & 1/6 & 1/2 & 4 \\
E2  & $102-107$ & 0 & 1/6 & 1/2 & 4 \\
E3  & $108-113$ & 0 & 1/6 & 1/2 & $-4$\\
E4  & $114-119$ & 0 & 1/6 & 1/2 & $-4$\\ \hline
Fa1 & $120-121$ & 0 & 0 & 1/2 & 0 \\
Fa2 & $122-127$ & 0 & 0 & 1/2 & 0\\ \hline
Fb1 & $128$     & 0 & 0 & 1/2 & 12 \\
Fb2 & $129-131$ & 0 & 0 & 1/2 & 4 \\
Fb3 & $132-134$ & 0 & 0 & 1/2 & $-4$ \\
Fb4 & $135$     & 0 & 0 & 1/2 & $-12$ \\ \hline
Fc1 & $136-141$ & 0 & 0 & 1/2 & 8 \\
Fc2 & $142-153$ & 0 & 0 & 1/2 & 0 \\
Fc3 & $154-159$ & 0 & 0 & 1/2 & $-8$\\ \hline
G1 & $160-165$  & 0 & 0 & 5/12 & 2 \\
G2 & $166-171$  & 0 & 0 & 5/12 & 2  \\
G3 & $172-183$  & 0 & 0 & 5/12 & 2  \\
G4 & $184-189$  & 0 & 0 & 5/12 & $-2$  \\
G5 & $190-201$  & 0 & 0 & 5/12 & $-2$  \\
G6 & $202-207$  & 0 & 0 & 5/12 & $-6$ \\ \hline
Ha & $208-213$  & 0 & 0 & 1/3 & 0  \\ \hline
Hb1 & $214-219$ & 0 & 0 & 1/3 & 4  \\
Hb2 & $220-225$ & 0 & 0 & 1/3 & 4 \\
Hb3 & $226-231$ & 0 & 0 & 1/3 & $-4$ \\
Hb4 & $232-237$ & 0 & 0 & 1/3 & $-4$ \\ \hline
I1  & $238-239$ & 0 & 0 & 1/4 & 6 \\
I2  & $240-245$ & 0 & 0 & 1/4 & $-2$\\
I3  & $246-251$ & 0 & 0 & 1/4 & 2\\
I4  & $252-253$ & 0 & 0 & 1/4 & $-6$\\ \hline
J1  &  $254$    & 0 & 0 & 0 & 0\\
J2  &  $255 $   & 0 & 0 & 0 & 0\\
\hline
\end{longtable}

\bibliography{sizeCitation}

\begin{thebibliography}{52}
\expandafter\ifx\csname natexlab\endcsname\relax\def\natexlab#1{#1}\fi
\expandafter\ifx\csname bibnamefont\endcsname\relax
  \def\bibnamefont#1{#1}\fi
\expandafter\ifx\csname bibfnamefont\endcsname\relax
  \def\bibfnamefont#1{#1}\fi
\expandafter\ifx\csname citenamefont\endcsname\relax
  \def\citenamefont#1{#1}\fi
\expandafter\ifx\csname url\endcsname\relax
  \def\url#1{\texttt{#1}}\fi
\expandafter\ifx\csname urlprefix\endcsname\relax\def\urlprefix{URL }\fi
\providecommand{\bibinfo}[2]{#2}
\providecommand{\eprint}[2][]{\url{#2}}

\bibitem[{\citenamefont{Kauffman}(1969)}]{K69}
\bibinfo{author}{\bibfnamefont{S.~A.} \bibnamefont{Kauffman}},
  \bibinfo{journal}{J. Theor. Biol.} \textbf{\bibinfo{volume}{22}},
  \bibinfo{pages}{437} (\bibinfo{year}{1969}).

\bibitem[{\citenamefont{Kauffman}(1993)}]{K93}
\bibinfo{author}{\bibfnamefont{S.~A.} \bibnamefont{Kauffman}},
  \emph{\bibinfo{title}{The origins of order}} (\bibinfo{publisher}{Oxford
  University Press}, \bibinfo{address}{New York}, \bibinfo{year}{1993}).

\bibitem[{\citenamefont{Aldana et~al.}(2003)\citenamefont{Aldana, Coppersmith,
  and Kadanoff}}]{ACK02}
\bibinfo{author}{\bibfnamefont{M.}~\bibnamefont{Aldana}},
  \bibinfo{author}{\bibfnamefont{S.}~\bibnamefont{Coppersmith}},
  \bibnamefont{and} \bibinfo{author}{\bibfnamefont{L.~P.}
  \bibnamefont{Kadanoff}}, in \emph{\bibinfo{booktitle}{Perspectives and
  Problems in Nonlinear Science}}, edited by
  \bibinfo{editor}{\bibfnamefont{E.}~\bibnamefont{Kaplan}},
  \bibinfo{editor}{\bibfnamefont{J.}~\bibnamefont{Marsden}}, \bibnamefont{and}
  \bibinfo{editor}{\bibfnamefont{K.}~\bibnamefont{Sreenivasan}}
  (\bibinfo{publisher}{Springer-Verleg}, \bibinfo{address}{Berlin},
  \bibinfo{year}{2003}).

\bibitem[{\citenamefont{Drossel}(2008)}]{Drossel07}
\bibinfo{author}{\bibfnamefont{B.}~\bibnamefont{Drossel}}, in
  \emph{\bibinfo{booktitle}{Reviews of Nonlinear Dynamics and Complexity (to be
  published)}}, edited by \bibinfo{editor}{\bibfnamefont{H.~G.}
  \bibnamefont{Schuster}} (\bibinfo{publisher}{Wiley-VCH},
  \bibinfo{address}{Berlin}, \bibinfo{year}{2008}), \eprint{arXiv:0706.3351}.

\bibitem[{\citenamefont{Bornholdt}(2005)}]{lessmore}
\bibinfo{author}{\bibfnamefont{S.}~\bibnamefont{Bornholdt}},
  \bibinfo{journal}{Science} \textbf{\bibinfo{volume}{310}},
  \bibinfo{pages}{449} (\bibinfo{year}{2005}).

\bibitem[{\citenamefont{Albert and Othmer}(2003)}]{lessE1}
\bibinfo{author}{\bibfnamefont{R.}~\bibnamefont{Albert}} \bibnamefont{and}
  \bibinfo{author}{\bibfnamefont{H.~G.} \bibnamefont{Othmer}},
  \bibinfo{journal}{J. Theor. Biol.} \textbf{\bibinfo{volume}{223}},
  \bibinfo{pages}{1} (\bibinfo{year}{2003}).

\bibitem[{\citenamefont{Li et~al.}(2004)\citenamefont{Li, Long, Lu, Quyang, and
  Tang}}]{lessE2}
\bibinfo{author}{\bibfnamefont{F.}~\bibnamefont{Li}},
  \bibinfo{author}{\bibfnamefont{T.}~\bibnamefont{Long}},
  \bibinfo{author}{\bibfnamefont{Y.}~\bibnamefont{Lu}},
  \bibinfo{author}{\bibfnamefont{Q.}~\bibnamefont{Quyang}}, \bibnamefont{and}
  \bibinfo{author}{\bibfnamefont{C.}~\bibnamefont{Tang}},
  \bibinfo{journal}{Proc. Natl. Acad. Sci. U.S.A.}
  \textbf{\bibinfo{volume}{101}}, \bibinfo{pages}{4781} (\bibinfo{year}{2004}).

\bibitem[{\citenamefont{Shmulevich et~al.}(2005)\citenamefont{Shmulevich,
  Kauffman, and Aldana}}]{SKA05}
\bibinfo{author}{\bibfnamefont{I.}~\bibnamefont{Shmulevich}},
  \bibinfo{author}{\bibfnamefont{S.~A.} \bibnamefont{Kauffman}},
  \bibnamefont{and} \bibinfo{author}{\bibfnamefont{M.}~\bibnamefont{Aldana}},
  \bibinfo{journal}{Proc. Natl. Acad. Sci. U.S.A.}
  \textbf{\bibinfo{volume}{102}}, \bibinfo{pages}{13439}
  (\bibinfo{year}{2005}).

\bibitem[{\citenamefont{Lau et~al.}(2007)\citenamefont{Lau, Ganguli, and
  Tang}}]{LGT07}
\bibinfo{author}{\bibfnamefont{K.~Y.} \bibnamefont{Lau}},
  \bibinfo{author}{\bibfnamefont{S.}~\bibnamefont{Ganguli}}, \bibnamefont{and}
  \bibinfo{author}{\bibfnamefont{C.}~\bibnamefont{Tang}},
  \bibinfo{journal}{Phys. Rev. E} \textbf{\bibinfo{volume}{75}},
  \bibinfo{pages}{051907} (\bibinfo{year}{2007}).

\bibitem[{\citenamefont{Kauffman and Smith}(1986)}]{KS86}
\bibinfo{author}{\bibfnamefont{S.~A.} \bibnamefont{Kauffman}} \bibnamefont{and}
  \bibinfo{author}{\bibfnamefont{R.~G.} \bibnamefont{Smith}},
  \bibinfo{journal}{Physica D} \textbf{\bibinfo{volume}{22}},
  \bibinfo{pages}{68} (\bibinfo{year}{1986}).

\bibitem[{\citenamefont{Bornholdt and Sneppen}(1998)}]{BS97}
\bibinfo{author}{\bibfnamefont{S.}~\bibnamefont{Bornholdt}} \bibnamefont{and}
  \bibinfo{author}{\bibfnamefont{K.}~\bibnamefont{Sneppen}},
  \bibinfo{journal}{Phys. Rev. Lett.} \textbf{\bibinfo{volume}{81}},
  \bibinfo{pages}{236} (\bibinfo{year}{1998}).

\bibitem[{\citenamefont{Paczuski et~al.}(2000)\citenamefont{Paczuski, Bassler,
  and Corral}}]{PBC00}
\bibinfo{author}{\bibfnamefont{M.}~\bibnamefont{Paczuski}},
  \bibinfo{author}{\bibfnamefont{K.~E.} \bibnamefont{Bassler}},
  \bibnamefont{and} \bibinfo{author}{\bibfnamefont{A.}~\bibnamefont{Corral}},
  \bibinfo{journal}{Phys. Rev. Lett.} \textbf{\bibinfo{volume}{84}},
  \bibinfo{pages}{3185} (\bibinfo{year}{2000}).

\bibitem[{\citenamefont{Bornholdt and Rohlf}(2000)}]{Bornholdt}
\bibinfo{author}{\bibfnamefont{S.}~\bibnamefont{Bornholdt}} \bibnamefont{and}
  \bibinfo{author}{\bibfnamefont{T.}~\bibnamefont{Rohlf}},
  \bibinfo{journal}{Phys. Rev. Lett.} \textbf{\bibinfo{volume}{84}},
  \bibinfo{pages}{6114} (\bibinfo{year}{2000}).

\bibitem[{\citenamefont{Bornholdt and Sneppen}(2000)}]{BS00}
\bibinfo{author}{\bibfnamefont{S.}~\bibnamefont{Bornholdt}} \bibnamefont{and}
  \bibinfo{author}{\bibfnamefont{K.}~\bibnamefont{Sneppen}},
  \bibinfo{journal}{Proc. R. Soc. Lond. B} \textbf{\bibinfo{volume}{267}},
  \bibinfo{pages}{2281} (\bibinfo{year}{2000}).

\bibitem[{\citenamefont{Luque et~al.}(2001)\citenamefont{Luque, Ballesteros,
  and Muro}}]{SOCRBN}
\bibinfo{author}{\bibfnamefont{B.}~\bibnamefont{Luque}},
  \bibinfo{author}{\bibfnamefont{F.~J.} \bibnamefont{Ballesteros}},
  \bibnamefont{and} \bibinfo{author}{\bibfnamefont{E.~M.} \bibnamefont{Muro}},
  \bibinfo{journal}{Phys. Rev. E} \textbf{\bibinfo{volume}{63}},
  \bibinfo{pages}{051913} (\bibinfo{year}{2001}).

\bibitem[{\citenamefont{Bassler et~al.}(2004)\citenamefont{Bassler, Lee, and
  Lee}}]{canalization}
\bibinfo{author}{\bibfnamefont{K.~E.} \bibnamefont{Bassler}},
  \bibinfo{author}{\bibfnamefont{C.}~\bibnamefont{Lee}}, \bibnamefont{and}
  \bibinfo{author}{\bibfnamefont{Y.}~\bibnamefont{Lee}},
  \bibinfo{journal}{Phys. Rev. Lett.} \textbf{\bibinfo{volume}{93}},
  \bibinfo{pages}{038101} (\bibinfo{year}{2004}).

\bibitem[{\citenamefont{Bassler and Liu}(2005)}]{canalization05}
\bibinfo{author}{\bibfnamefont{K.~E.} \bibnamefont{Bassler}} \bibnamefont{and}
  \bibinfo{author}{\bibfnamefont{M.}~\bibnamefont{Liu}}, in
  \emph{\bibinfo{booktitle}{Noise in Complex Systems and Stochastic Dynamics
  III, Proc. of SPIE Vol. 5845}}, edited by
  \bibinfo{editor}{\bibfnamefont{L.~B.} \bibnamefont{Kish}},
  \bibinfo{editor}{\bibfnamefont{K.}~\bibnamefont{Lindenberg}},
  \bibnamefont{and} \bibinfo{editor}{\bibfnamefont{Z.}~\bibnamefont{Gingl}}
  (\bibinfo{publisher}{SPIE}, \bibinfo{address}{Bellingham, WA},
  \bibinfo{year}{2005}).

\bibitem[{\citenamefont{Liu and Bassler}(2006)}]{coevolution}
\bibinfo{author}{\bibfnamefont{M.}~\bibnamefont{Liu}} \bibnamefont{and}
  \bibinfo{author}{\bibfnamefont{K.~E.} \bibnamefont{Bassler}},
  \bibinfo{journal}{Phys. Rev. E} \textbf{\bibinfo{volume}{74}},
  \bibinfo{pages}{041910} (\bibinfo{year}{2006}).

\bibitem[{\citenamefont{Szejka and Drossel}(2007)}]{SD07}
\bibinfo{author}{\bibfnamefont{A.}~\bibnamefont{Szejka}} \bibnamefont{and}
  \bibinfo{author}{\bibfnamefont{B.}~\bibnamefont{Drossel}},
  \bibinfo{journal}{Eur. Phys. J. B} \textbf{\bibinfo{volume}{56}},
  \bibinfo{pages}{373} (\bibinfo{year}{2007}).

\bibitem[{\citenamefont{Rohlf}(2007)}]{R07}
\bibinfo{author}{\bibfnamefont{T.}~\bibnamefont{Rohlf}} (\bibinfo{year}{2007}),
  \eprint{arXiv:0708.1637}.

\bibitem[{\citenamefont{Braunewell and Bornholdt}(2007)}]{BB07}
\bibinfo{author}{\bibfnamefont{S.}~\bibnamefont{Braunewell}} \bibnamefont{and}
  \bibinfo{author}{\bibfnamefont{S.}~\bibnamefont{Bornholdt}}
  (\bibinfo{year}{2007}), \eprint{arXiv:0707.1407}.

\bibitem[{\citenamefont{Challet and Zhang}(1997)}]{CZ97}
\bibinfo{author}{\bibfnamefont{D.}~\bibnamefont{Challet}} \bibnamefont{and}
  \bibinfo{author}{\bibfnamefont{Y.~C.} \bibnamefont{Zhang}},
  \bibinfo{journal}{Physica A} \textbf{\bibinfo{volume}{246}},
  \bibinfo{pages}{407} (\bibinfo{year}{1997}).

\bibitem[{\citenamefont{Reichhardt and Bassler}(2007)}]{symmetry}
\bibinfo{author}{\bibfnamefont{C.~J.~O.} \bibnamefont{Reichhardt}}
  \bibnamefont{and} \bibinfo{author}{\bibfnamefont{K.~E.}
  \bibnamefont{Bassler}}, \bibinfo{journal}{J. Phys. A: Math. Theor.}
  \textbf{\bibinfo{volume}{40}}, \bibinfo{pages}{4339} (\bibinfo{year}{2007}).

\bibitem[{\citenamefont{Waddington}(1942)}]{W42}
\bibinfo{author}{\bibfnamefont{C.}~\bibnamefont{Waddington}},
  \bibinfo{journal}{Nature} \textbf{\bibinfo{volume}{150}},
  \bibinfo{pages}{563} (\bibinfo{year}{1942}).

\bibitem[{\citenamefont{Rutherford and Lindquist}(1998)}]{RL98}
\bibinfo{author}{\bibfnamefont{R.}~\bibnamefont{Rutherford}} \bibnamefont{and}
  \bibinfo{author}{\bibfnamefont{S.}~\bibnamefont{Lindquist}},
  \bibinfo{journal}{Nature} \textbf{\bibinfo{volume}{396}},
  \bibinfo{pages}{336} (\bibinfo{year}{1998}).

\bibitem[{\citenamefont{Quietsch et~al.}(2002)\citenamefont{Quietsch, Sangster,
  and Lindquist}}]{Q02}
\bibinfo{author}{\bibfnamefont{C.}~\bibnamefont{Quietsch}},
  \bibinfo{author}{\bibfnamefont{T.~A.} \bibnamefont{Sangster}},
  \bibnamefont{and}
  \bibinfo{author}{\bibfnamefont{S.}~\bibnamefont{Lindquist}},
  \bibinfo{journal}{Nature} \textbf{\bibinfo{volume}{417}},
  \bibinfo{pages}{618} (\bibinfo{year}{2002}).

\bibitem[{\citenamefont{Bergman and Siegal}(2003)}]{BS03}
\bibinfo{author}{\bibfnamefont{A.}~\bibnamefont{Bergman}} \bibnamefont{and}
  \bibinfo{author}{\bibfnamefont{M.}~\bibnamefont{Siegal}},
  \bibinfo{journal}{Nature} \textbf{\bibinfo{volume}{424}},
  \bibinfo{pages}{549} (\bibinfo{year}{2003}).

\bibitem[{\citenamefont{Wagner}(2005)}]{W05}
\bibinfo{author}{\bibfnamefont{A.}~\bibnamefont{Wagner}},
  \emph{\bibinfo{title}{Robustness and Evolvability in living systems}}
  (\bibinfo{publisher}{Princeton University Press}, \bibinfo{address}{New
  Jersey}, \bibinfo{year}{2005}).

\bibitem[{\citenamefont{Harrison}(1963)}]{Harrison}
\bibinfo{author}{\bibfnamefont{M.~A.} \bibnamefont{Harrison}},
  \bibinfo{journal}{SIAM} \textbf{\bibinfo{volume}{11}}, \bibinfo{pages}{806}
  (\bibinfo{year}{1963}).

\bibitem[{\citenamefont{Derrida and Pomeau}(1986)}]{DP86}
\bibinfo{author}{\bibfnamefont{B.}~\bibnamefont{Derrida}} \bibnamefont{and}
  \bibinfo{author}{\bibfnamefont{Y.}~\bibnamefont{Pomeau}},
  \bibinfo{journal}{EuroPhys. Lett.} \textbf{\bibinfo{volume}{1}},
  \bibinfo{pages}{45} (\bibinfo{year}{1986}).

\bibitem[{\citenamefont{Derrida and Weisbuch}(1986)}]{DW86}
\bibinfo{author}{\bibfnamefont{B.}~\bibnamefont{Derrida}} \bibnamefont{and}
  \bibinfo{author}{\bibfnamefont{G.}~\bibnamefont{Weisbuch}},
  \bibinfo{journal}{J. Phys.} \textbf{\bibinfo{volume}{47}},
  \bibinfo{pages}{1297} (\bibinfo{year}{1986}).

\bibitem[{\citenamefont{Flyvbjerg}(1988)}]{F88}
\bibinfo{author}{\bibfnamefont{H.}~\bibnamefont{Flyvbjerg}},
  \bibinfo{journal}{J. Phys. A} \textbf{\bibinfo{volume}{21}},
  \bibinfo{pages}{L955} (\bibinfo{year}{1988}).

\bibitem[{\citenamefont{Bastolla and Parisi}(1997)}]{BP97}
\bibinfo{author}{\bibfnamefont{U.}~\bibnamefont{Bastolla}} \bibnamefont{and}
  \bibinfo{author}{\bibfnamefont{G.}~\bibnamefont{Parisi}},
  \bibinfo{journal}{J. Theor. Biol} \textbf{\bibinfo{volume}{187}},
  \bibinfo{pages}{117} (\bibinfo{year}{1997}).

\bibitem[{\citenamefont{Bastolla and Parisi}(1998)}]{BP98}
\bibinfo{author}{\bibfnamefont{U.}~\bibnamefont{Bastolla}} \bibnamefont{and}
  \bibinfo{author}{\bibfnamefont{G.}~\bibnamefont{Parisi}},
  \bibinfo{journal}{Physica D} \textbf{\bibinfo{volume}{115}},
  \bibinfo{pages}{203} (\bibinfo{year}{1998}).

\bibitem[{\citenamefont{Fox and Hill}(2001)}]{Fox}
\bibinfo{author}{\bibfnamefont{J.~J.} \bibnamefont{Fox}} \bibnamefont{and}
  \bibinfo{author}{\bibfnamefont{C.~C.} \bibnamefont{Hill}},
  \bibinfo{journal}{Chaos} \textbf{\bibinfo{volume}{11}}, \bibinfo{pages}{809}
  (\bibinfo{year}{2001}).

\bibitem[{\citenamefont{Aldana}(2003)}]{Aldana1}
\bibinfo{author}{\bibfnamefont{M.}~\bibnamefont{Aldana}},
  \bibinfo{journal}{Physica D} \textbf{\bibinfo{volume}{185}},
  \bibinfo{pages}{45} (\bibinfo{year}{2003}).

\bibitem[{\citenamefont{Aldana and Cluzel}(2003)}]{Aldana2}
\bibinfo{author}{\bibfnamefont{M.}~\bibnamefont{Aldana}} \bibnamefont{and}
  \bibinfo{author}{\bibfnamefont{P.}~\bibnamefont{Cluzel}},
  \bibinfo{journal}{Proc. Natl. Acad. Sci. U.S.A}
  \textbf{\bibinfo{volume}{100}}, \bibinfo{pages}{8710} (\bibinfo{year}{2003}).

\bibitem[{\citenamefont{Socolar and Kauffman}(2003)}]{Socolar03}
\bibinfo{author}{\bibfnamefont{J.~E.~S.} \bibnamefont{Socolar}}
  \bibnamefont{and} \bibinfo{author}{\bibfnamefont{S.~A.}
  \bibnamefont{Kauffman}}, \bibinfo{journal}{Phys. Rev. Lett.}
  \textbf{\bibinfo{volume}{90}}, \bibinfo{pages}{068702}
  (\bibinfo{year}{2003}).

\bibitem[{\citenamefont{Samuelsson and Troein}(2003)}]{ST03}
\bibinfo{author}{\bibfnamefont{B.}~\bibnamefont{Samuelsson}} \bibnamefont{and}
  \bibinfo{author}{\bibfnamefont{C.}~\bibnamefont{Troein}},
  \bibinfo{journal}{Phys. Rev. Lett.} \textbf{\bibinfo{volume}{90}},
  \bibinfo{pages}{098701} (\bibinfo{year}{2003}).

\bibitem[{\citenamefont{Kaufman et~al.}(2005)\citenamefont{Kaufman, Mihaljev,
  and Drossel}}]{drossel1}
\bibinfo{author}{\bibfnamefont{V.}~\bibnamefont{Kaufman}},
  \bibinfo{author}{\bibfnamefont{T.}~\bibnamefont{Mihaljev}}, \bibnamefont{and}
  \bibinfo{author}{\bibfnamefont{B.}~\bibnamefont{Drossel}},
  \bibinfo{journal}{Phys. Rev. E} \textbf{\bibinfo{volume}{72}},
  \bibinfo{pages}{046124} (\bibinfo{year}{2005}).

\bibitem[{\citenamefont{Greil and Drossel}(2005)}]{drossel2}
\bibinfo{author}{\bibfnamefont{F.}~\bibnamefont{Greil}} \bibnamefont{and}
  \bibinfo{author}{\bibfnamefont{B.}~\bibnamefont{Drossel}},
  \bibinfo{journal}{Phys. Rev. Lett.} \textbf{\bibinfo{volume}{95}},
  \bibinfo{pages}{048701} (\bibinfo{year}{2005}).

\bibitem[{\citenamefont{Drossel}(2005)}]{drossel3}
\bibinfo{author}{\bibfnamefont{B.}~\bibnamefont{Drossel}},
  \bibinfo{journal}{Phys. Rev. E} \textbf{\bibinfo{volume}{72}},
  \bibinfo{pages}{016110} (\bibinfo{year}{2005}).

\bibitem[{\citenamefont{Drossel et~al.}(2005)\citenamefont{Drossel, Mihaljev,
  and Greil}}]{drossel4}
\bibinfo{author}{\bibfnamefont{B.}~\bibnamefont{Drossel}},
  \bibinfo{author}{\bibfnamefont{T.}~\bibnamefont{Mihaljev}}, \bibnamefont{and}
  \bibinfo{author}{\bibfnamefont{F.}~\bibnamefont{Greil}},
  \bibinfo{journal}{Phys. Rev. Lett.} \textbf{\bibinfo{volume}{94}},
  \bibinfo{pages}{088701} (\bibinfo{year}{2005}).

\bibitem[{\citenamefont{Moreira and Amaral}(2005)}]{MA05}
\bibinfo{author}{\bibfnamefont{A.~A.} \bibnamefont{Moreira}} \bibnamefont{and}
  \bibinfo{author}{\bibfnamefont{L.~A.~N.} \bibnamefont{Amaral}},
  \bibinfo{journal}{Phys. Rev. Lett.} \textbf{\bibinfo{volume}{94}},
  \bibinfo{pages}{218702} (\bibinfo{year}{2005}).

\bibitem[{\citenamefont{Arthur}(1994)}]{A94}
\bibinfo{author}{\bibfnamefont{W.~B.} \bibnamefont{Arthur}},
  \bibinfo{journal}{Am. Econ. Rev.} \textbf{\bibinfo{volume}{84}},
  \bibinfo{pages}{406} (\bibinfo{year}{1994}).

\bibitem[{\citenamefont{Bak and Sneppen}(1993)}]{BS93}
\bibinfo{author}{\bibfnamefont{P.}~\bibnamefont{Bak}} \bibnamefont{and}
  \bibinfo{author}{\bibfnamefont{K.}~\bibnamefont{Sneppen}},
  \bibinfo{journal}{Phys. Rev. Lett.} \textbf{\bibinfo{volume}{71}},
  \bibinfo{pages}{4083} (\bibinfo{year}{1993}).

\bibitem[{\citenamefont{P{\'o}lya}(1937)}]{polya37}
\bibinfo{author}{\bibfnamefont{G.}~\bibnamefont{P{\'o}lya}},
  \bibinfo{journal}{Acta. Math.} \textbf{\bibinfo{volume}{38}},
  \bibinfo{pages}{145} (\bibinfo{year}{1937}).

\bibitem[{\citenamefont{Lo et~al.}(2000)\citenamefont{Lo, Hui, and
  Johnson}}]{LHJ00}
\bibinfo{author}{\bibfnamefont{T.~S.} \bibnamefont{Lo}},
  \bibinfo{author}{\bibfnamefont{P.~M.} \bibnamefont{Hui}}, \bibnamefont{and}
  \bibinfo{author}{\bibfnamefont{N.~F.} \bibnamefont{Johnson}},
  \bibinfo{journal}{Phys. Rev. E} \textbf{\bibinfo{volume}{62}},
  \bibinfo{pages}{4393} (\bibinfo{year}{2000}).

\bibitem[{\citenamefont{Hart et~al.}(2000)\citenamefont{Hart, Jefferies, Hui,
  and Johnson}}]{HJHJ00}
\bibinfo{author}{\bibfnamefont{M.}~\bibnamefont{Hart}},
  \bibinfo{author}{\bibfnamefont{P.}~\bibnamefont{Jefferies}},
  \bibinfo{author}{\bibfnamefont{P.~M.} \bibnamefont{Hui}}, \bibnamefont{and}
  \bibinfo{author}{\bibfnamefont{N.~F.} \bibnamefont{Johnson}},
  \bibinfo{journal}{Eur. Phys. J. B} \textbf{\bibinfo{volume}{20}},
  \bibinfo{pages}{547} (\bibinfo{year}{2000}).

\bibitem[{\citenamefont{Anghel et~al.}(2004)\citenamefont{Anghel, Toroczkai,
  Bassler, and Korniss}}]{ATBK04}
\bibinfo{author}{\bibfnamefont{M.}~\bibnamefont{Anghel}},
  \bibinfo{author}{\bibfnamefont{Z.}~\bibnamefont{Toroczkai}},
  \bibinfo{author}{\bibfnamefont{K.~E.} \bibnamefont{Bassler}},
  \bibnamefont{and} \bibinfo{author}{\bibfnamefont{G.}~\bibnamefont{Korniss}},
  \bibinfo{journal}{Phys. Rev. Lett.} \textbf{\bibinfo{volume}{92}},
  \bibinfo{pages}{058701} (\bibinfo{year}{2004}).

\bibitem[{\citenamefont{Milo et~al.}(2002)\citenamefont{Milo, Orr, Itzkovitz,
  Kashtan, Chklovskii, and Alon}}]{UriAlon}
\bibinfo{author}{\bibfnamefont{R.}~\bibnamefont{Milo}},
  \bibinfo{author}{\bibfnamefont{S.~S.} \bibnamefont{Orr}},
  \bibinfo{author}{\bibfnamefont{S.}~\bibnamefont{Itzkovitz}},
  \bibinfo{author}{\bibfnamefont{N.}~\bibnamefont{Kashtan}},
  \bibinfo{author}{\bibfnamefont{D.}~\bibnamefont{Chklovskii}},
  \bibnamefont{and} \bibinfo{author}{\bibfnamefont{U.}~\bibnamefont{Alon}},
  \bibinfo{journal}{Science} \textbf{\bibinfo{volume}{298}},
  \bibinfo{pages}{824} (\bibinfo{year}{2002}).

\bibitem[{\citenamefont{Shen-Orr et~al.}(2002)\citenamefont{Shen-Orr, Milo,
  Mangan, and Alon}}]{SMMA02}
\bibinfo{author}{\bibfnamefont{S.}~\bibnamefont{Shen-Orr}},
  \bibinfo{author}{\bibfnamefont{R.}~\bibnamefont{Milo}},
  \bibinfo{author}{\bibfnamefont{S.}~\bibnamefont{Mangan}}, \bibnamefont{and}
  \bibinfo{author}{\bibfnamefont{U.}~\bibnamefont{Alon}},
  \bibinfo{journal}{Nature Genetics} \textbf{\bibinfo{volume}{31}},
  \bibinfo{pages}{64} (\bibinfo{year}{2002}).

\end{thebibliography}

\end{document}